\documentclass[11pt]{elsarticle}

\pdfoutput=1

\makeatletter 
\def\ps@pprintTitle{%
  \let\@oddhead\@empty
  \let\@evenhead\@empty
  \let\@oddfoot\@empty
  \let\@evenfoot\@oddfoot
}
\makeatother

\usepackage{url}
\usepackage{breakurl}
\usepackage[breaklinks,
            colorlinks = true,
            linkcolor = blue,
            urlcolor  = blue,
            citecolor = blue,
            anchorcolor = blue]{hyperref}

\usepackage{lineno}

\usepackage{graphicx}
\usepackage[margin=1.25in]{geometry}
\usepackage[usenames,dvipsnames]{color}

\newcommand\acp{\begin{center}
\rule[-0.2in]{\hsize}{0.01in}\\\rule{\hsize}{0.01in}\\
\vskip 0.1in Submitted to the  Proceedings\\ 
of the African Conference on Fundamental and Applied Physics
    \vskip 0.05in
    {\it Second Edition, ACP2021, March 7--11, 2022 --- Virtual Event}\\
\rule{\hsize}{0.01in}\\\rule[+0.2in]{\hsize}{0.01in} \\
\end{center}}

\usepackage[firstpage=true]{background}
\backgroundsetup{contents={\parbox{6.5in}{\acp}}, scale=1,placement=top,opacity=1,color=black,position={3.25in,1.2in}}

\usepackage{caption}
\usepackage{subcaption}
\usepackage{fancyhdr}
\fancypagestyle{plain}{%
  \fancyhf{}%
  \fancyhead[C]{}
  \fancyfoot[C]{\thepage}
}

\fancypagestyle{empty}{%
  \fancyhf{}%
  \fancyhead[C]{{\it ACP2021, March 7--11, 2022 --- Virtual Edition}}
  \fancyfoot[C]{\thepage}
}
\begin{document}

\begin{frontmatter}


\title{Measurement of Low-Activity Uranium Contamination using Bayesian Statistical Decision Theory}

\author[add1]{Hanan Arahmane\corref{cor1}}
\ead{hanan.arahmane@gmail.com}
\author[add2]{Jonathan Dumazert}
\author[add1]{Eric Barat}
\author[add1]{Thomas Dautremer}
\author[add1]{Frédérick Carrel}
\author[add1]{Nicolas Dufour}
\author[add1]{Maugan Michel}

\cortext[cor1]{Corresponding Author}

\address[add1]{Université Paris-Saclay, CEA, List, F-91120 Palaiseau, France}
\address[add2]{CEA-DAM, DIF, F-91297 Arpajon, France}

\begin{abstract}
\noindent 
Amongst the various technical challenges in the field of radiation detection is the need to carry out accurate low-level radioactivity measurements in the presence of large fluctuations in the natural radiation background, while lowering the false alarm rates. Several studies, using statistical inference, have been proposed to overcome this challenge. This work presents an advanced statistical approach for decision-making in the field of nuclear decommissioning. The results indicate that the proposed method allows to adjust the confidence degree in the stationarity of the background signal. It also ensures an acceptable tradeoff between the True Detection Rate (TDR), the False Alarm Rate (FAR) and the response time, and is consistent with the user’s requirements. 
\end{abstract}


\end{frontmatter}

%



\section{Introduction}
\label{sec:intro}
\noindent
Accurate detection of radioactivity in low count rates is crucial for many practical radiation-detection applications. The particular approach considered here is that of statistical analysis that has recently provided new horizons in the area of radioactive detection. It is one of the reliable and accurate analysis for determining the most effective detection strategy with reasonable operational requirements. In this paper, we aim to measure a low-activity uranium contamination on concrete surfaces, with varying enrichment encountered levels within a basic nuclear facility. Our surface contamination limit was taken as $ S_A = 500~Bq/0.1~m^2$~\cite{arahmane2021reliable}. In this context, we have developed an advanced Bayesian method based on a Kibble bivariate gamma distribution, referred to as the Bayesian Kibble Dirichlet (BKD) test. It allows to adjust the confidence degree in the stationarity of the background signal. In this perspective, the BKD test combines the properties of an absolute and a relative hypothesis statistical test, that were both developed with the same application at hand~\cite{arahmane2021reliable}, and respectively referred to as the Bayesian Absolute test by mixture of Multivariate Poisson laws (BAM) and the Bayesian Relative test by mixture of Multinomial laws (BRM). Note that unlike the BKD test, the modeling of the BAM test was based on the univariate gamma distribution. The implementation of the Bayesian approach were based on \textit{a priori} vectors constructed from the coupling of experimental data acquired within a basic nuclear facility using a high-purity germanium detector (HPGe), as well as simulated data with Monte Carlo N-Particles 6 transport code (MCNP6.1)~\cite{werner2017mcnp}. The performance evaluation and characterization of the Bayesian method were performed using classical receiver operating characteristic curves (ROC)~\cite{metz1978basic} with the study of the background signal variations effect.

\section{Materials and Methods}
\label{sec:mater}
\subsection{Uranium gamma peaks of interest and background signal}
\noindent
 Detection of low levels of uranium contamination is to determine its presence by means of an HPGe detector. To do so, we used the main gamma-ray emission lines from the decay chains of U-235 and U-238~\cite{arahmane2021reliable}: (i) from the U-235 chain, one main line at 185.7 keV, and an additional triplet at {143.8; 163.4; 205.3} keV, and (ii) from the U-238 chain, one main line at 1001.0 keV. 
 
Regarding the background signal, it should be noted that the detection of a low-surface activity uranium contamination on concrete surfaces, based on gamma-ray emission, can be disturbed by the presence of the background signatures, in particular K-40, and the decay chains of U-238 and Th-232. In order to quantify this variation within nuclear facility that is subdivided into three subunits (SUs), we performed the acquisitions over source-detector distance $ d= 22.36~cm $  for a surface of $ 0.1~ m^2 $. Then, we calculated the relative deviation between two separate spots inside each SU.

\subsection{Simulation of the detector response}
\noindent
The performance evaluation of the Bayesian method needs prior knowledge of the expected spectral response from a uranium contamination corresponding to the minimum surface activities to be detected (MSAD) (i.e., decommissioning criterion). MSAD used in this study is equivalent to $500~Bq/0.1 m^2/2\pi$. To this end, we followed three steps. First, we constructed a transfer function relating the emission rates $(s^{-1})$ with the five gamma lines identified above at this MSAD. Second,  we built a numerical model of the HPGe detector using the MCNP6.1, in order to estimate the spectral count rate $ (s^{-1})$ expected at its output from MSAD and enrichment level in U-235.  Finally, we simulated the U-235 enrichment levels of the spectral response of the detector as well. It should be noted that each of the above stated SU has average expected U-235 enrichment levels that were set to be $ 0.62 $ wt \%$,2.03 $ wt \%$,$ and$~6.84 $ wt \% $ $. These enrichment levels are likely to fluctuate around these average values. As a result, we used regular U-235 enrichment mass fractions corresponding to the natural ones ($\approx 0.7 $ wt \%$), 3 $ wt \%$, $ and $ 8 $ wt \%$ $ for SU1, SU2, and SU3 respectively, in order to construct tested spectra.

\subsection	{Bayesian Methodology}
\noindent
The construction of Bayesian statistical tests, either BAM, BRM and BKD needs some prior Knowledge. The spectral signatures of both the background signal, and an expected uranium contamination inside a nuclear facility, are the priors retained in this study. We recall here that the prior spectra of the background signal, and of an expected uranium contamination are respectively provided by in-situ acquisitions, and Monte Carlo simulations based on MCNP6.1.

The construction of the Bayesian test leading to the acceptance of one of the two hypotheses $H_0$ or $H_1$ makes use of Bayes' theorem as follows:
\begin{equation}\label{my_first_eqn}
\frac{P(H_0\mid\mathbf{m_{ref},m_{test} )}}{P(H_1\mid\mathbf{ m_{ref},m_{test}} )} = \frac{P\mathbf{(m_{ref},m_{test}}\mid H_0 )P(H_0 )}{P\mathbf{(m_{ref},m_{test}}\mid H_1 )P(H_1)}
\end{equation}

where $ m_{ref}$ and $m_{test} $ the multichannel countings, $ P(H_0)$ and $ P(H_1)$ are priors on the respective likelihoods of hypotheses $H_0$ and $H_1$.

Without any \textit{a priori} information of the presence of a uranium contamination, these probabilities are taken equal by default: $P(H_0 ) $ = $P(H_1) $ = $ 0.5 $. Moreover, we note that: $P(H_1\mid\mathbf{ m_{ref},m_{test}}) $ = $1-P(H_0\mid\mathbf{ m_{ref},m_{test}} )$. As a result, by taking into account these equalities in equation (\ref{my_first_eqn}) the posterior probability of hypothesis $ H_0$, that will be ultimately compared to a decision threshold, can be expressed as: 

\begin{equation}
{P(H_0\mid\mathbf{m_{ref},m_{test} )}} = \frac{1}{1 + \kappa _{BAM,BRM,BKD}}
\end{equation}

with the Bayes factor expressed as:  
\begin{equation}
\kappa _{BAM,BRM,BKD} = \frac{P\mathbf{(m_{ref},m_{test}}\mid H_1 )}{P\mathbf{(m_{ref},m_{test}}\mid H_0)}
\end{equation}

From the last three equations, we can see that the construction of the Bayesian tests is the comparison of $P(H_0\mid\mathbf{ m_{ref},m_{test}})$ and $P(H_1\mid\mathbf{ m_{ref},m_{test}} )$ through the computation of the Bayes factor $\kappa _{BAM,BRM,BKD}$.

It should be noted that the process of adjusting the properties of the BKD test in order to retrieve the limiting behaviors of the absolute (BAM) and relative (BRM) tests, relies essentially upon the parametrization of the BKD test with respect to the correlation coefficient of Kibble law $\rho$~\cite{balakrishna2009bivariate}. Therefore, the implementation of the BKD test requires the setting of a parameter $\rho$ which governs the behavior of BKD between that of BRM ($\rho$ = $0$) and that of BAM ($\rho$ →$1$). 

\section{Results and Conclusions}
\begin{figure}[!h]
  \begin{subfigure}{0.5\textwidth}
\includegraphics[width=\textwidth]{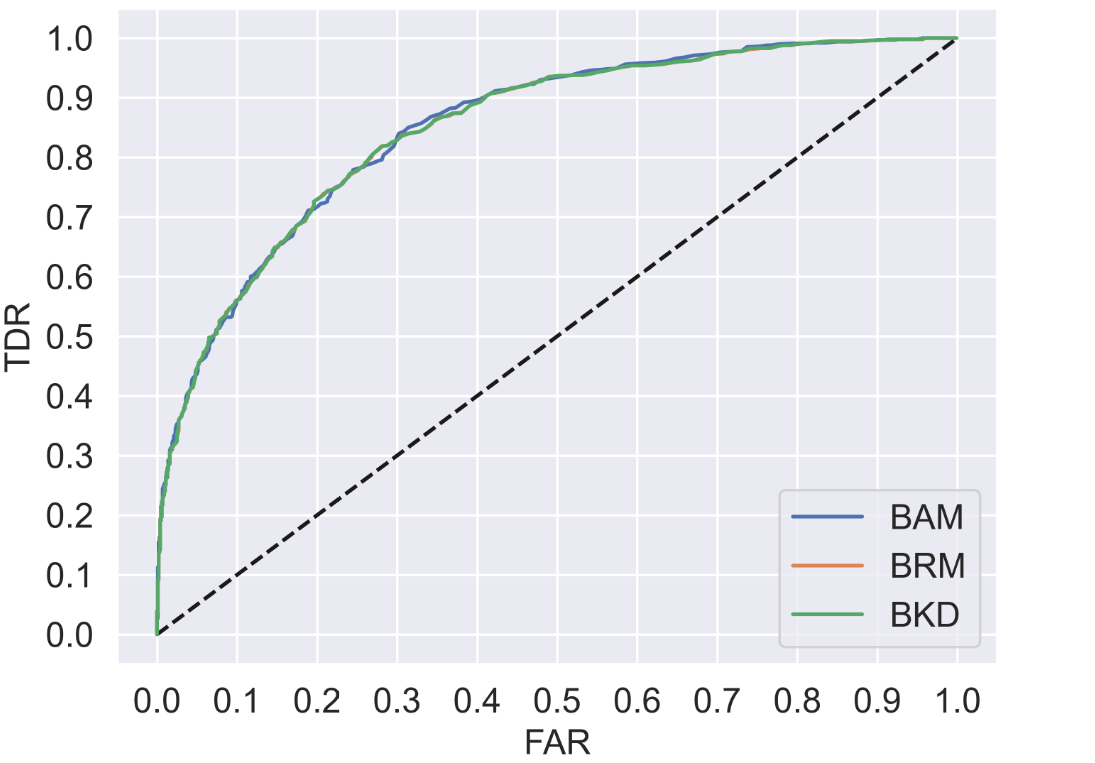}
\vspace{-5mm}
    \caption{$\rho$ = $0$} \label{fig:ROCstaa}
  \end{subfigure}%
  \hfill   
  \begin{subfigure}{0.5\textwidth}
   \includegraphics[width=\textwidth]{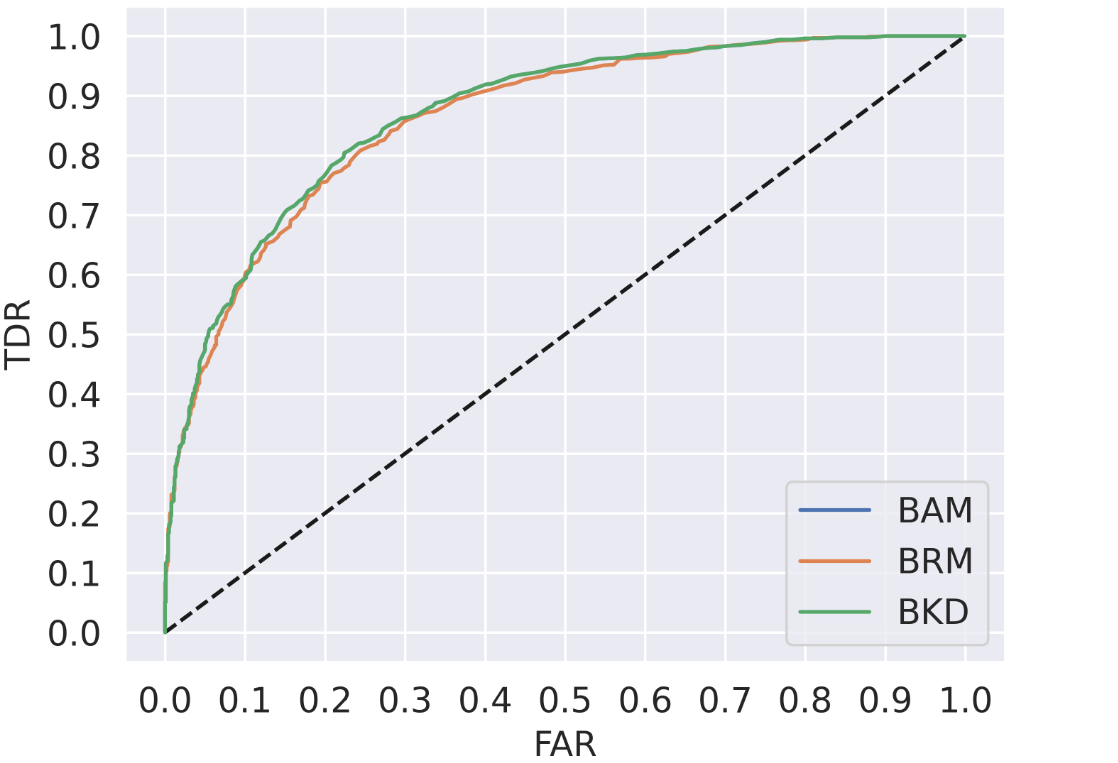}
   \vspace{-5mm}
    \caption{$\rho$ → $1$} \label{fig:ROCstab}
  \end{subfigure}%
  \vspace{-3mm}
\caption{ROC curves for an enrichment level of $0.7$ wt\%$ $, an integration time $t$ = $3000$ s, a stationary background signal, and $\rho$ : (a) $\rho$ = $0$, (b) $\rho$ → $1$.}
\label{fig:ROCsta}
\end{figure}

\begin{figure}[!h]
  \begin{subfigure}{0.5\textwidth}
\includegraphics[width=\textwidth]{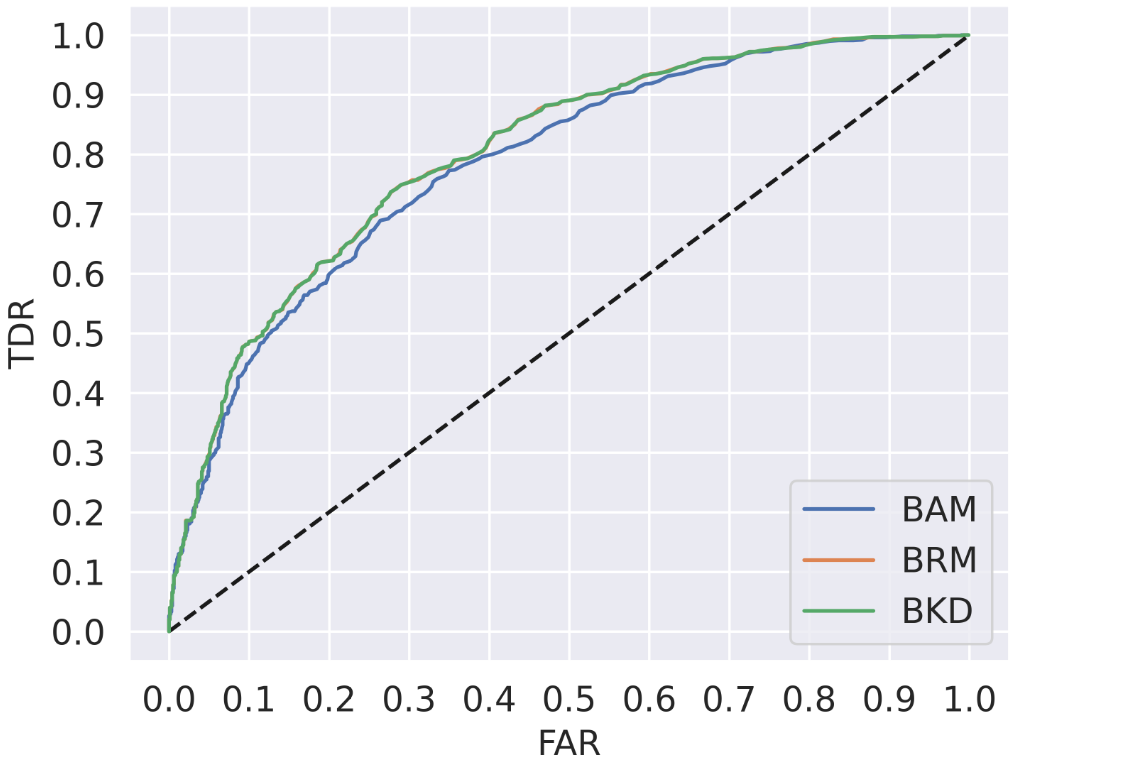}
\vspace{-5mm}
    \caption{$\rho$ = $0$} \label{fig:ROCnstaa}
  \end{subfigure}%
  \hfill   
  \begin{subfigure}{0.5\textwidth}
   \includegraphics[width=\textwidth]{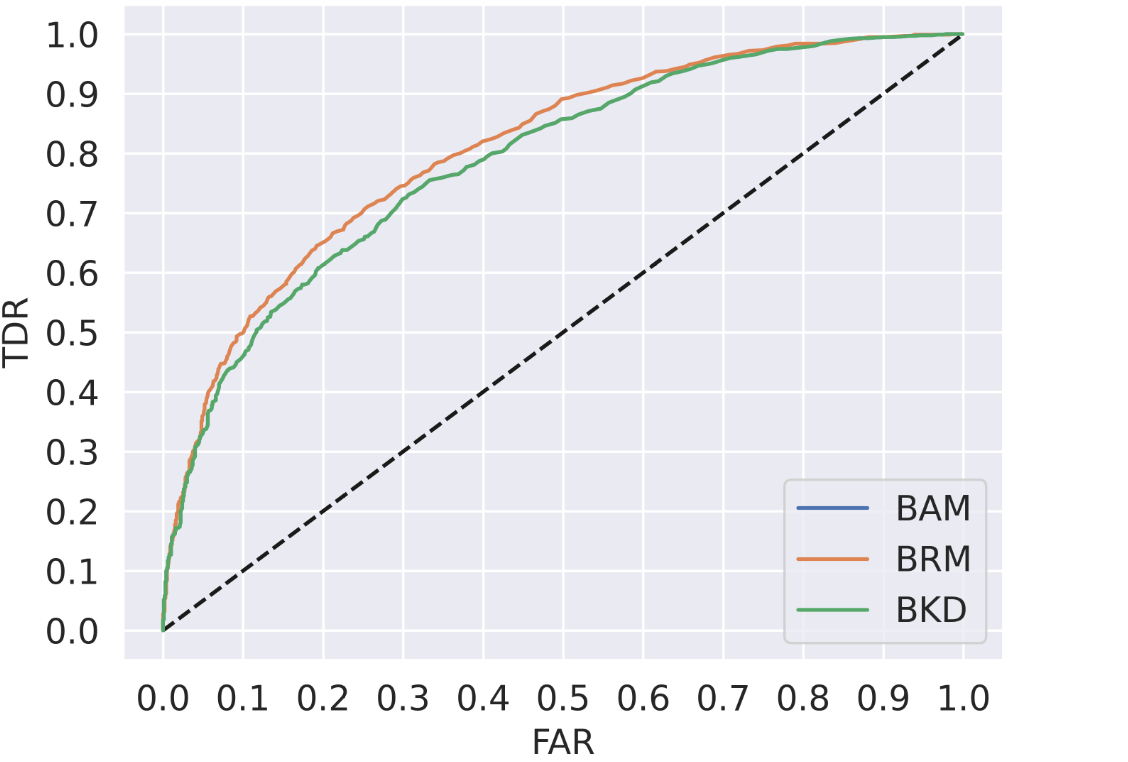}
   \vspace{-5mm}
    \caption{$\rho$ → $1$} \label{fig:ROCnstab}
  \end{subfigure}%
  \vspace{-3mm}
\caption{ROC curves for an enrichment level of $0.7$ wt\%$ $, an integration time $t$ = $3000$ s, a non-stationary background signal, and $\rho$ : (c) $\rho$ = $0$, (d) $\rho$ → $1$.}
\label{fig:ROCnsta}
\end{figure}

\noindent
In this section, we present the comparative performance study of the BKD test with the absolute (BAM) and relative (BRM) ones by using a ROC-based analysis. The scenarios involving these three settings of $\rho$ in both cases of a stationary or non-stationary background signal, in amplitude as well as in shape, yield the ROC curves in Figures~\ref{fig:ROCsta} and~\ref{fig:ROCnsta}.

These ROCs  allow to identify explicit trends. From the figures, we can see that the BKD test, whether the background signal is stationary and non-stationary, has the same behavior as the relative BRM test (Figures~\ref{fig:ROCstaa} and~\ref{fig:ROCnstaa}) in the case of $\rho$ = $0$, while it presents the properties of the absolute BAM test in the case of $\rho$ →$1$ (Figures~\ref{fig:ROCstab} and~\ref{fig:ROCnstab}). Furthermore, the results showed that a reliable detection of uranium was reachable with a significantly higher TDR/FAR tradeoff. However, we can observe that the variation of the background signal, affected the detection performance of the Bayesian tests, which is normal, due to an increase in the statistical fluctuations in the spectra.

\bibliographystyle{elsarticle-num}
\bibliography{myreferences} 

\begin{thebibliography}{1}
\expandafter\ifx\csname url\endcsname\relax
  \def\url#1{\texttt{#1}}\fi
\expandafter\ifx\csname urlprefix\endcsname\relax\def\urlprefix{URL }\fi
\expandafter\ifx\csname href\endcsname\relax
  \def\href#1#2{#2} \def\path#1{#1}\fi

\bibitem{ACP2021-report}
{K\'et\'evi A. Assamagan, Obinna Abah, Amare Abebe, Stephen Avery, et al.},
  {Activity report of the Second African Conference on Fundamental and Applied
  Physics, ACP2021}, {arXiv:2204.01882} ({2022}).
\newblock \href {https://doi.org/https://doi.org/10.48550/arXiv.2204.01882}
  {\path{doi:https://doi.org/10.48550/arXiv.2204.01882}}.

\bibitem{ASP-COVID}
{Kossi Amouzouvi, Kétévi A. Assamagan, Somiéalo Azote, Simon H. Connell,
  Jean Baptiste Fankam Fankam, Fenosoa Fanomezana, Aluwani Guga, Cyrille E.
  Haliya, Toivo S. Mabote, Francisco Fenias Macucule, Dephney Mathebula,
  Azwinndini Muronga, Kondwani C. C. Mwale, Ann Njeri, Ebode F. Onyie, Laza
  Rakotondravohitra, George Zimba}, {A model of COVID-19 pandemic evolution in
  African countries}, Scientific African vol. 14 ({2021}) e00987.
\newblock \href {http://arxiv.org/abs/2104.09675} {\path{arXiv:2104.09675}},
  \href {https://doi.org/https://doi.org/10.1016/j.sciaf.2021.e00987}
  {\path{doi:https://doi.org/10.1016/j.sciaf.2021.e00987}}.

\bibitem{asfap}
{K\'et\'evi A. Assamagan, Simon H. Connell, Farida Fassi, Fairouz Malek,
  Shaaban I. Khalil, et al.}, {The African Strategy for Fundamental and Applied
  Physics}, \url{https://africanphysicsstrategy.org/} (2021).

\end{thebibliography}


\begin{thebibliography}{10}
\expandafter\ifx\csname url\endcsname\relax
  \def\url#1{\texttt{#1}}\fi
\expandafter\ifx\csname urlprefix\endcsname\relax\def\urlprefix{URL }\fi
\expandafter\ifx\csname href\endcsname\relax
  \def\href#1#2{#2} \def\path#1{#1}\fi

\bibitem{ASP2021-reports}
{Kétévi A. Assamagan, Bobby Acharya, Temitope Adenuga, Mohamed Chabab,
  Kenneth Cecire, Simon H. Connell, Anne E. Dabrowski, Christine Darve, Farida
  Fassi, Jonathan R. Ellis, Fernando Ferroni, Mounia Laassiri, Steve G.
  Muanza}, {Activity report of the African School of Physics, 2019-2021},
  {arXiv:2109.00509} ({2019--2021}).
\newblock \href {https://doi.org/https://doi.org/10.48550/arXiv.2109.00509}
  {\path{doi:https://doi.org/10.48550/arXiv.2109.00509}}.

\bibitem{ASP}
{B. S. Acharya, K. A. Assamagan, A. E. Dabrowski, C. Darve, J. Ellis, S.
  Muanza}, {The African School of Physics},
  \url{https://www.africanschoolofphysics.org/}.

\bibitem{ASP-reports}
{Activity reports of African School of Physics},
  \url{http://africanschoolofphysics.web.cern.ch/2010/asp2010.pdf,
  https://africanschoolofphysics.web.cern.ch/asp2012/asp2012_final.pdf,
  https://www.africanschoolofphysics.org/wp-content/uploads/2014/11/asp2014.pdf,
  https://www.africanschoolofphysics.org/wp-content/uploads/2019/08/ASP2016-FinalReport.pdf,
  https://www.africanschoolofphysics.org/wp-content/uploads/2019/08/ASP2018.pdf}
  (2010-2018).

\bibitem{ASP-COVID}
{Kossi Amouzouvi, Kétévi A. Assamagan, Somiéalo Azote, Simon H. Connell,
  Jean Baptiste Fankam Fankam, Fenosoa Fanomezana, Aluwani Guga, Cyrille E.
  Haliya, Toivo S. Mabote, Francisco Fenias Macucule, Dephney Mathebula,
  Azwinndini Muronga, Kondwani C. C. Mwale, Ann Njeri, Ebode F. Onyie, Laza
  Rakotondravohitra, George Zimba}, {A model of COVID-19 pandemic evolution in
  African countries}, Scientific African vol. 14 ({2021}) e00987.
\newblock \href {http://arxiv.org/abs/2104.09675} {\path{arXiv:2104.09675}},
  \href {https://doi.org/https://doi.org/10.1016/j.sciaf.2021.e00987}
  {\path{doi:https://doi.org/10.1016/j.sciaf.2021.e00987}}.

\bibitem{asp2018}
{B. S. Acharya, K. A. Assamagan, M. Backes, K. Cecire, A. E. Dabrowski, C.
  Darve, J. Ellis, J. A. Gray, E. Kasai, S. Muanza, J. Ndjamba1, A. Philander,
  M. Shahungu, G. Simon, D. Singh, R. Steenkamp, R. Voss, A. Zulu}, {Activity
  Report on the Fifth Biennial African School of Fundamental Physics and
  Applications},
  \url{https://www.africanschoolofphysics.org/wp-content/uploads/2019/08/ASP2018.pdf}
  (2018).

\bibitem{acp2021}
{K\'et\'evi A. Assamagan, Mohamed Chabab, Farida Fassi, Ulrich Goerlach,
  Mohamed Gouighri, et al.}, {The second African Conference on Fundamental and
  Applied Physics}, \url{https://indico.cern.ch/event/1060503/} (2022).

\bibitem{marrakesh}
{Cadi Ayyad University}, \url{https://www.uca.ma/}.

\bibitem{rabat}
{Mohammed V University}, \url{http://www.um5.ac.ma/um5/}.

\bibitem{AfPS}
{The African Physical Society}, \url{https://www.africanphysicalsociety.org/}.

\bibitem{asfap}
{K\'et\'evi A. Assamagan, Simon H. Connell, Farida Fassi, Fairouz Malek,
  Shaaban I. Khalil, et al.}, {The African Strategy for Fundamental and Applied
  Physics}, \url{https://africanphysicsstrategy.org/} (2021).

\bibitem{aas}
{The African Academy of Sciences}, \url{https://www.aasciences.africa/}.

\bibitem{unesco}
{The United Nations Educational, Scientific and Cultural Organization},
  \url{https://en.unesco.org/}.

\bibitem{iaea}
{International Atomic Energy Agency}, \url{https://www.iaea.org/}.

\bibitem{cambridge}
{University of Cambridge}, \url{https://www.cam.ac.uk/}.

\bibitem{psi}
{Paul Scherrer Institute}, \url{https://www.psi.ch/en}.

\bibitem{desy}
{Deutsches Elektronen-Synchrotron}, \url{https://www.desy.de/}.

\bibitem{ucad}
{Universit\'e Cheikh Anta Diop de Dakar}, \url{https://www.ucad.sn/}.

\bibitem{ICTP}
{The International Center for Theoretical Physics}, \url{https://www.ictp.it/}.

\bibitem{IIP}
{Investing In People (IIP) ASBL}, \url{www.semainedelasciencerdc.org}.

\end{thebibliography}


\begin{thebibliography}{1}
\expandafter\ifx\csname url\endcsname\relax
  \def\url#1{\texttt{#1}}\fi
\expandafter\ifx\csname urlprefix\endcsname\relax\def\urlprefix{URL }\fi
\expandafter\ifx\csname href\endcsname\relax
  \def\href#1#2{#2} \def\path#1{#1}\fi

\bibitem{arahmane2021reliable}
H.~Arahmane, J.~Dumazert, E.~Barat, T.~Dautremer, N.~Dufour, F.~Carrel,
  M.~Michel, F.~Lain{\'e}, A reliable absolute and relative bayesian method for
  nuclear decommissioning: Low-level radioactivity detection with gamma-ray
  spectrometry, IEEE Transactions on Instrumentation and Measurement 70 (2021)
  1--18.

\bibitem{werner2017mcnp}
C.~J. Werner, et~al., Mcnp users manual-code version 6.2, Los Alamos national
  laboratory (2017).

\bibitem{metz1978basic}
C.~E. Metz, Basic principles of roc analysis, in: Seminars in nuclear medicine,
  Vol.~8, Elsevier, 1978, pp. 283--298.

\bibitem{balakrishna2009bivariate}
N.~Balakrishna, C.~D. Lai, Bivariate gamma and related distributions, in:
  Continuous Bivariate Distributions, Springer, 2009, pp. 305--350.

\end{thebibliography}

\end{document}